\documentclass{emulateapj}

\newcommand{\cii}{\mbox{\rm [\ion{C}{2}]}}

\newcommand{\jone}{\mbox{($1\rightarrow0$)}}

\newcommand{\fscii}{($^2$P$_{3/2}\rightarrow^2$P$_{1/2}$)}
\newcommand{\percmcu}{cm$^{-3}$}

\newcommand{\kmpers}{\mbox{km~s$^{-1}$}}

\newcommand{\Ico}{\mbox{\rm I$_{\rm CO}$}}
\newcommand{\Lco}{\mbox{\rm L$_{\rm CO}$}}
\newcommand{\Lsun}{\mbox{\rm L$_{\odot}$}}

\newcommand{\Icii}{\mbox{\rm I$_{\rm [CII]}$}}
\newcommand{\Lcii}{\mbox{\rm L$_{\rm [CII]}$}}
\newcommand{\Fcii}{\mbox{\rm F$_{\rm [CII]}$}}
\newcommand{\Ifir}{\mbox{\rm I$_{\rm FIR}$}}
\newcommand{\Lfir}{\mbox{\rm L$_{\rm FIR}$}}
\newcommand{\Ffir}{\mbox{\rm F$_{\rm FIR}$}}

\begin{document}
\title{The Faintness of the 158 $\mu$m [\ion{C}{2}] Transition \\
in the z=6.42 Quasar SDSS J1148+5251}
\author{Alberto D. Bolatto}
\affil{Department of Astronomy \& Radio Astronomy Laboratory,\\ 
University of California at Berkeley, \\
601 Campbell Hall, Berkeley, CA 94720-3411, USA}
\author{James Di Francesco \& Chris J. Willott}
\affil{National Research Council of Canada, Herzberg Institute of 
Astrophysics, \\
5071 West Saanich Road, Victoria, BC, V9E 2E7, CANADA}

\begin{abstract}
We report the non-detection of the \cii\ \fscii\ 157.74 $\mu$m
transition in the $z=6.42$ quasar SDSS J1148+5251 after 37.5 hours of
integration with the James Clerk Maxwell Telescope.  This transition
is the main cooling line of the star-forming interstellar medium, and
usually the brightest FIR line in galaxies. Our observed 1 $\sigma$
RMS = 1.3 mK in the $T_{A}^{*}$ scale translates to $\Lcii<2.6 \times
10^{9}$ L$_\odot$.  Using a recent estimate of the far-infrared
continuum of this quasar, we derive for SDSS J1148+5251
$\Lcii/\Lfir<5\times10^{-4}$, a ratio similar to that observed in
local ultra-luminous infrared galaxies but considerably smaller than
what is typical in nearby normal and starburst galaxies. This
indicates that the small \Lcii/\Lfir\ ratio observed locally in
luminous far-infrared objects also persists at the highest redshifts.
\end{abstract}

\keywords{quasars: individual (J114816.64+525150.3) --- quasars:
emission lines --- early universe --- radio lines: galaxies}

\section{Introduction}

The most distant quasar in the universe known at the time of this
writing is SDSS J114816.64+525150.3 (hereafter SDSS J1148+5251) at a
redshift $z=6.42$ (Fan et al. 2003; Walter et al. 2003; Bertoldi et
al. 2003a). With a bolometric luminosity of L$_{\rm bol}\sim10^{14}$
L$_\odot$, this is an extremely luminous object powered by a
$\sim3\times10^{9}$ M$_\odot$ supermassive black hole at its core
(Willott et al. 2003).  The high far-infrared (FIR) luminosity of this
object implies that the host galaxy is forming stars at the prodigious
rate of 3000 M$_\odot$ yr$^{-1}$ (Bertoldi et al. 2003b). At this
redshift the universe is only 840 million years old and both the black
hole mass and star formation rate imply the existence of a very
massive galaxy formed from one of the rarest high-density peaks in the
matter distribution. Although there are hints that SDSS J1148+5251 may
be weakly gravitationally lensed (White et al. 2003), we assume no
lensing magnification in this paper.

The recent detection of bright molecular and FIR continuum emission
from the host galaxy of SDSS J1148+5251 (Walter et al. 2003; Bertoldi
et al. 2003a; Bertoldi et al. 2003b) prompted us to observe the \cii\
\fscii\ 157.74 $\mu$m fine structure transition in this object. This
is the main cooling transition of the star-forming interstellar medium
(e.g., Tielens \& Hollenbach 1985), and is commonly the single
brightest emission line in galaxies.  For example, Stacey et
al. (1991) found that most galaxies emit 0.1\% to 1\% of their
far-infrared luminosities in this line alone.

Given its extremely high luminosity and its relationship to star
formation activity, the redshifted \cii\ 158 $\mu$m transition is
attractive as a star formation indicator at high-$z$ (e.g., Stark
1997).  This transition is conveniently placed into the atmospheric
1~mm window for redshifts $z\gtrsim6.2$, making it accessible to
ground-based instrumentation.  Furthermore, a spectroscopic approach
to detecting high-$z$ sources has the clear advantage of containing
redshift information, unlike continuum observations, thus avoiding
problems of source confusion that stem from the low angular resolution
of single-dish radio telescopes equipped with bolometer arrays.  As
early as 1997, however, Infrared Space Observatory (ISO) observations
suggested potential problems with this method.  Indeed, Malhotra et
al. (1997) and Luhman et al. (1998) found that the proportionality
between the \cii\ and FIR continuum luminosities observed in nearby
star forming galaxies broke down for luminous infrared galaxies, where
the \cii\ luminosity appeared not to exceed $\Lcii\sim10^9$ \Lsun.
These conclusions, confirmed and expanded in a recent analysis of ISO
data by Luhman et al. (2003), have since cast doubts on the usefulness
of redshifted \cii\ to probe star formation in the distant
universe. The observations presented here have bearing on this matter:
does the paucity of \cii\ emission observed in local ultra-luminous
infrared galaxies (ULIRGs) hold for objects in the early universe?

\section{Observations and Results}

We observed the quasar SDSS J1148+5251 using the James Clerk Maxwell
Telescope (JCMT) at Mauna Kea, Hawaii\footnote{The JCMT is operated 
by the Joint Astronomy Centre in Hilo, Hawaii on behalf of the parent 
organizations Particle Physics and Astronomy Research Council in the 
United Kingdom, the National Research Council of Canada and The 
Netherlands Organization for Scientific Research.}.  The observations 
were performed in the flexible queue mode over several shifts between
2003 November 15 and 2004 January 30, using 8~Hz beam-position switching 
over 60 second cycles with offsets of 60\arcsec\ in azimuth.  Standard
calibration observations, which include pointing and focus, were
performed at the beginning and sometimes near the end of each shift.
The RxA3 dual sideband heterodyne receiver of the JCMT was tuned to
place 256.172 GHz at the center of the lower sideband, which assumes a
rest frame wavelength for the \cii\ transition of $\lambda=157.74$
$\mu$m and a redshift of $z=6.419$ as reported for the CO detections
of this object (Walter et al. 2003; Bertoldi et al. 2003a). We
utilized the 1840 MHz bandpass of the Digital Autocorrelation
Spectrometer which allowed us to cover a velocity range approximately
$\pm1100$ \kmpers\ around the expected emission, equivalent to a 
redshift range $z\approx6.39-6.44$ (by comparison, the detected CO 
lines are $\sim300$ \kmpers\ wide).  Approximately 48 hours of 
integration were devoted to this project.

Individual spectra obtained from 10 minute-long integrations were read
into SPECX, merged, and converted to FITS format. The data were then
read into COMB where spectra showing abnormal RMS (due to baseline
problems) were expunged, and good quality individual integrations were
then combined using the built-in weighting scheme (i.e.,
$t_{int}/T^2_{sys}$), then Hanning-smoothed to 10~MHz wide
channels. The final spectrum is shown in Figure 1, the result of 37.5
hours of integration after the removal of a linear baseline. Its RMS
is 1.3~mK in the central 1000 \kmpers\ region, or 32 mJy assuming an
aperture efficiency $\approx$ 0.56 for the JCMT at the observing
frequency.  The integrated intensity of the \cii\ line is
$\Icii\approx -1.7\pm1.9$ Jy~\kmpers\ $\approx
(-1.4\pm1.6)\times10^{-17}$ erg s$^{-1}$ cm$^{-2}$.  The calibration
uncertainty of these observations could be as large as $30\%$ due to
the poorly characterized sideband ratio of RxA3 at 256 GHz combined
with the complex frequency structure of the receiver noise
temperature.

To convert this measurement to a luminosity in a manner consistent
with other recent work on this source, we adopt the WMAP cosmology
with H$_0=71$ \kmpers, $\Omega_{\rm m}=0.27$, and
$\Omega_\Lambda=0.73$ (Spergel et al. 2003). The luminosity distance
to a source at a redshift $z=6.419$ is then
$D_l\approx1.97\times10^{29}$~cm. Consequently, the \cii\ luminosity
of SDSS J1148+5251 is $\Lcii<2\times10^{9}$ L$_\odot$.  Using the FIR
luminosity $\Lfir\approx1.2\times10^{13}$ L$_\odot$ determined by
Bertoldi et al. (2003b) from 1~mm continuum observations carried out
with the MAMBO array, the \cii/FIR ratio is
$\Lcii/\Lfir<1.7\times10^{-4}$. Taking into account that changing the
assumed 40~K dust temperature by $\pm10$~K will change the derived
\Lfir\ by a factor of 2 (Bertoldi et al. 2003b) and the aforementioned
uncertainties in the \cii\ calibration, a more conservative limit is
$\Lcii/\Lfir<5\times10^{-4}$.

\section{Discussion}

The limit obtained on the \cii\ emission in J1148+5251 confirms some
of the suggested caveats on the usefulness of redshifted \cii\
emission as a beacon for star formation in the early universe.  The
\cii\ line in this very luminous FIR object is disproportionally
faint.  Indeed, our limit suggests a ratio in this quasar that appears
typical of the local sample of 15 ULIRGs studied by Luhman et al.
(2003). Their analysis found a median \cii/FIR flux ratio of
$\Fcii/\Ffir=3.9\times10^{-4}$ in ULIRGs whereas a median
$\Fcii/\Ffir=3.1\times10^{-3}$ was reported for a sample of 60 normal
star-forming galaxies by Malhotra et al. (2001).  Only one in five
ULIRGs approached the lower end of the distribution of \cii/FIR ratios
found in normal galaxies, with $\Fcii/\Ffir\sim10^{-3}$.  The \cii\
luminosities of the ULIRG systems studied by Luhman et al. had a mean
value $\Lcii\sim 1\times10^{9}$ L$_\odot$ with none exceeding $3.5
\times 10^9$ L$_\odot$. If the brightest \cii\ luminosities that can
be expected from objects at $z\sim6.5$ were similar to the
luminosities of local ULIRGs, observations that probe such regime need
to be $\sim3-10$ times more sensitive than what we have acheived --- a
feat probably beyond the current reach of single-dish instruments but
which may be attainable by the IRAM Plateau de Bure Interferometer or
certainly by the future Atacama Large Millimeter Array.

What are the reasons behind the deficit of \cii\ emission relative to
the FIR continuum? Malhotra et al. (1997; 2001) find a strong
correlation between decreasing \Lcii/\Lfir\ ratio and increasing
\Lfir\ and FIR color temperature, indicating a smooth transition
between the ratio observed in normal luminosity galaxies and that in
ULIRGs, which in turn suggests that the same mechanism is responsible
for low ratios in both types of objects.  Indeed, about 7\% of the
sources in the Malhotra et al. (2001) sample have measured ratios or
limits that are smaller than the $\Lcii/\Lfir\sim5\times10^{-4}$ more
typical of ULIRGs, although their luminosities are
$\Lfir\sim10^{10}-10^{11}$ \Lsun.  The physical conditions of the gas
in SDSS J1148+5251 may hold clues as to what mechanism is responsible
for the low observed ratios.  For example, we can place this object in
the \cii/FIR vs. CO/FIR plots obtained from photodissociation region
(PDR) models (e.g., see Kaufman et al. 1999) using our limit and the
CO observations and estimate some of its physical parameters. The
observational limit for the luminosity of the CO \jone\ transition in
SDSS J1148+5251 reported by Bertoldi et al.  (2003a) is
$\Lco<7\times10^6$ \Lsun, which yields $\Lco/\Lfir<6\times10^{-7}$.
Their multitransition LVG analysis, however, suggests that the
excitation conditions of the CO are very similar to those found by
Bradford et al.  (2003) in the starburst galaxy NGC~253.  The expected
CO \jone\ luminosity would then be close to the the observational
limit, i.e., $\Lco/\Lfir\sim6\times10^{-7}$.  Using Fig. 16 of Kaufman
et al. we can infer a UV radiation field strength $G_0\gtrsim600$ in
units of the local interstellar radiation field (Habing 1968), and
densities $n\gtrsim10^5$ \percmcu. In this determination the CO/FIR
ratio constrains the radiation field, while the \cii/FIR ratio
constrains the gas density.  A PDR with an incident radiation field
$G_0\gtrsim600$ would have a surface temperature $T\sim200$~K, rapidly
decreasing as the gas becomes increasingly molecular.  These
conditions are actually not peculiar, and are very similar to those
found in a variety of starburst galaxies and Galactic star-forming
regions (Stacey et al. 1991).

Whether or not a simple analysis like that described above can be
applied with impunity to a complex system such as an AGN is debatable.
Based on their available data, Malhotra et al. (1997; 2001) and Luhman et
al.  (2003) concluded that the observed deficit of \cii\ emission
relative to the continuum emission in FIR-luminous objects is
consistent with: 1) low \cii\ emission arising from PDRs with high
$G_0/n$, and/or 2) excess FIR emission not related to PDRs, but
perhaps arising from a enshrouded AGN component.  In the first case,
the conversion of UV photons into electrons that heat the gas is
inefficient because the dust grains rapidly become positively charged
and their photoelectric yield is dramatically reduced. Thus, the gas
temperature (and with it the \cii\ brightness) does not increase
proportionally with the radiation field but the dust temperature still
does, leading to low \Lcii/\Lfir\ ratios.  In the second case, the
\cii\ emission related to the AGN core is quenched for much the same
reasons, as well as because most of the carbon will likely be in a
higher ionization state.  Copious FIR radiation will be generated,
however, if the AGN is at least partially embedded in a dusty
envelope. Although this latter scenario naturally explains the deficit
of \cii\ emission in a quasar, we caution that there is a dearth of
additional evidence for it.  For example, one may expect that an
excess of FIR emission associated with dust near the ionized core of
the AGN will result in a lower than usual $\Lco/\Lfir$ ratio. The
$\Lco/\Lfir\sim6\times10^{-7}$ ratio inferred for SDSS J1148+5251,
however, is unremarkable when compared to a sample of local starburst
galaxies and Galactic star-forming regions (Stacey et al. 1991).
Indeed, this quasar falls into a region of the \Icii/\Ifir\ vs.
\Ico/\Ifir\ space mostly occupied by ULIRGs (see Fig. 8a of Genzel \&
Cesarsky 2000), showing that the observed ratios are not peculiarly
associated with AGN activity.  Nevertheless, without a detection of
the CO \jone\ transition or a stronger limit, it is difficult to
develop this analysis any further.

\section{Summary and Conclusion}

Our redshifted \cii\ observations at $\nu\sim256$~GHz have failed to
detect the fine structure 158 $\mu$m \cii\ transition in this high-$z$
quasar. We place a 1 $\sigma$ limit on the \cii\ luminosity of SDSS
J1148+5251 of $\Lcii<2.6\times10^9$ \Lsun\ (including 30\% calibration
uncertainty). Given the observed FIR flux of this source, this places
an upper limit on its \cii\ to FIR ratio of
$\Lcii/\Lfir<5\times10^{-4}$, substantially lower than what is found
in most local star-forming galaxies but similar to what is observed in
nearby ultra-luminous IR galaxies.  If the cause of this \cii\ deficit
in SDSS J1148+5251 was a large FIR component due to partial dust
reprocessing of the AGN radiation, thus unrelated to star formation
activity, the $\sim$3000 M$_\odot$ yr$^{-1}$ star formation rate
obtained from the FIR continuum could be a substantial overestimate.
Obtaining a better SED for the FIR continuum emission, to constrain
better the FIR luminosity, as well as improving the sensitivity of the
CO \jone\ observations, may help to elucidate this matter.

Will redshifted \cii\ observations open a new window onto high-$z$
galaxies?  Our results are based on observations of just one object at
high redshift. A larger sample of $z>6$ sources with high FIR
luminosities must be observed in the \cii\ transition to establish
whether low $\Lcii/\Lfir$ ratios are a general phenomenon.  If a
maximum luminosity $\Lcii\sim10^9$ \Lsun\ were confirmed for most
sources, the Atacama Large Millimeter Array would still be able to
detect the \cii\ line and spatially resolve its distribution and
kinematics. The answer, then, is a hopeful {\em yes}.

\acknowledgements
We thank Gerald Moriarty-Schieven, Ming Zhu, the JCMT Telescope
Support Specialists, and the numerous visiting JCMT observers who
obtained data for this project on our behalf. We would also like
to thank Fabian Walter, Gordon Stacey, and the anonymous referee.

\clearpage
\begin{figure}
\plotone{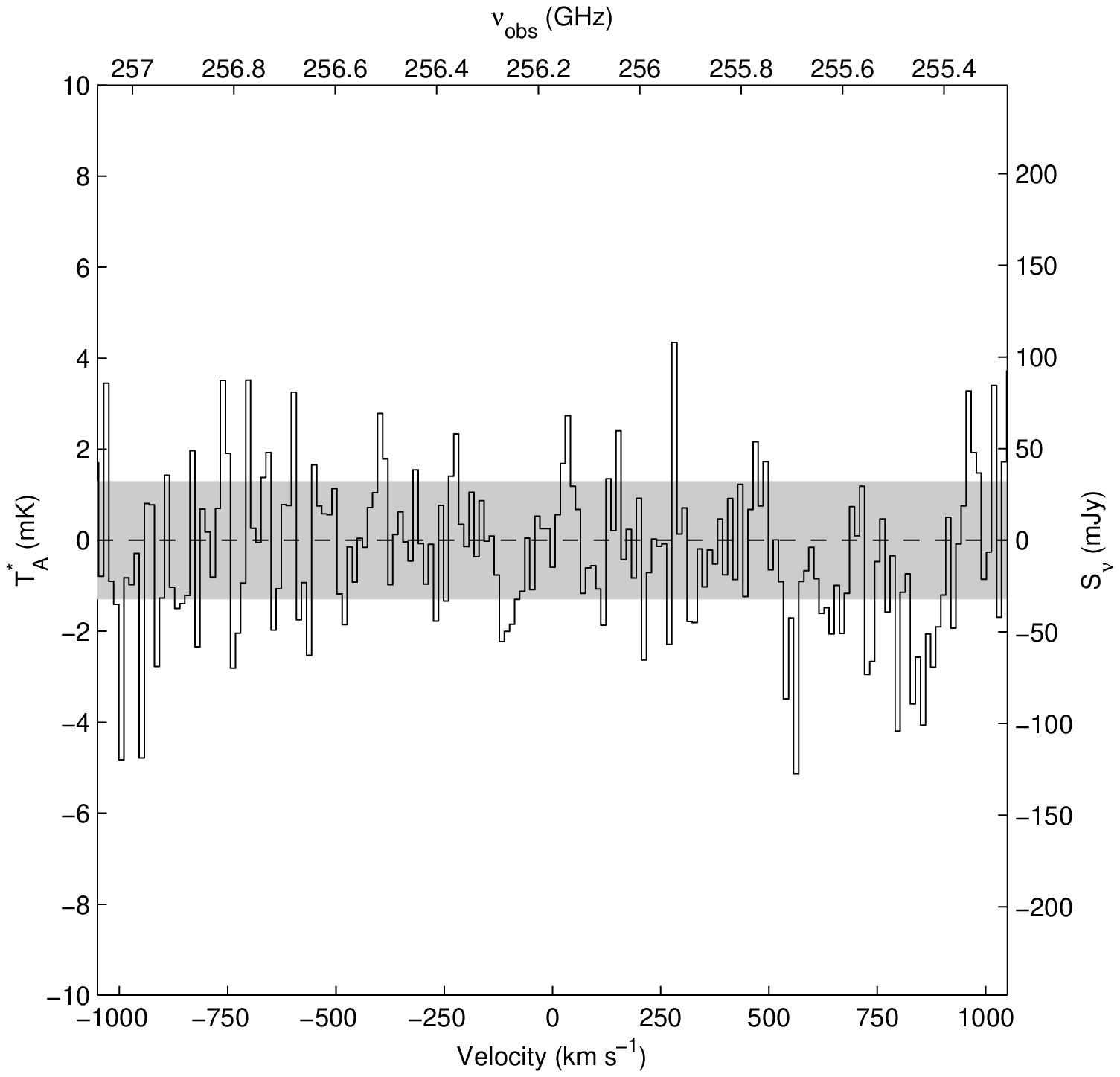}
\caption{The redshifted \cii\ spectrum obtained at the JCMT, with a
total of 37.5 hours of integration. The RMS in 10 MHz wide channels
($\sim11$ \kmpers) in the central 1000 \kmpers, illustrated by
the gray region, is 1.3 mK in the T$_A^*$ scale. This corresponds to
$\approx32$ mJy, using an aperture efficiency $\eta_{ap}\sim0.56$.}
\end{figure}
 
\end{document}